\documentstyle[12pt,aasms4]{article}
\def\puncspace{\ifmmode\,\else{\ifcat.\C{\if.\C\else\if,\C\else\if?\C\else%
\if:\C\else\if;\C\else\if-\C\else\if)\C\else\if/\C\else\if]\C\else\if'\C%
\else\space\fi\fi\fi\fi\fi\fi\fi\fi\fi\fi}%
\else\if\empty\C\else\if\space\C\else\space\fi\fi\fi}\fi}
\def\SP{\let\\=\empty\futurelet\C\puncspace }
\def\etal{et\SP al.\SP }

\def\deg{$^\circ$\ }
\def\kms{kms$^{-1}$\ }
 
\begin{document}
 
\title{STUDY OF A SLICE AT +9\deg TO +15\deg OF DECLINATION: I. THE NEUTRAL 
HYDROGEN CONTENT OF GALAXIES IN LOOSE GROUPS\altaffilmark{1}}

\author{M.A.G. Maia and C.N.A. Willmer} 
\affil{ Electronic mail: maia@on.br, cnaw@on.br} 
\affil{Departamento de Astronomia, Observat\'orio Nacional, Rua
General Jos\'e Cristino 77, Rio de Janeiro, 20921-030, Brazil}

\author{L.N. da Costa} 
\affil{European Southern Observatory, Karl-Schwarzschild-Str. 2, D-85748 
Garching b. M\"unchen, Germany and Departamento de Astronomia, Observat\'orio 
Nacional, Rua General Jos\'e Cristino 77, Rio de Janeiro, 20921-030, Brazil }

\altaffiltext{1}{Research reported herein is based partly on observations 
obtained at Arecibo Observatory (Puerto Rico), Complejo Astronomico El Leoncito 
(Argentina), Laborat\'orio Nacional de Astrof\'\i sica (Brazil) and  European 
Southern Observatory (Chile).}

\begin{abstract}
We examine the \ion{H}{1} content of spiral galaxies in groups by using a
catalog of loose groups of galaxies identified in a magnitude limited 
sample ($m_{Z} \le 15.7$) spanning the range $8^h \le \alpha \le 18^h$ in right 
ascension and $+9$\deg $\le \delta \le +15$\deg in declination.  
The redshift completeness of the galaxy sample is $\sim$ 95\%.  
No significant effect of \ion{H}{1} depletion is found, although there may be 
a hint that the earliest type spirals are slightly deficient.

\end{abstract}


\clearpage
 
\section{INTRODUCTION}

Evidence that the internal properties of galaxies are in some ways determined by 
the large scale clustering has accumulated throughout the years.  One of the first 
results suggesting this was presented by Davis \& Geller (1976) who showed that 
early-type galaxies have different clustering properties than late-types.  
By studying a sample of rich clusters, Dressler (1980) was able to show that a
well-defined relation exists over five orders of magnitude in density between the
local density of galaxies and the relative proportions of different morphological 
types, the so called morphology-density relation.   
This was later shown to extend to lower density regions such as groups of galaxies 
by Postman \& Geller (1984), and Maia \& da Costa (1990).  All these works show 
that the fraction of early-type galaxies increase with local density. 

Several other galaxy properties  have been claimed to be affected by the
environment, such as the formation of cD galaxies, enhancement of star formation rate, 
presence of active galactic nuclei, bars in spiral galaxies, colors, and far infra-red 
emission, among others. 
In particular, it has been found that the neutral Hydrogen (\ion{H}{1}) content 
of galaxies of a given morphological type depends on the environment (e.g., 
Haynes \& Giovanelli 1986; Magri \etal 1988; Huchtmeier \& Richter 1989; 
Scodeggio \& Gavazzi 1993; and more recently Maia \etal 1994) in the sense that 
galaxies in denser regions, such as the core of clusters of galaxies, are 
\ion{H}{1} deficient when compared to ``field'' galaxies (e.g., Haynes \& Giovanelli 
1986).  This \ion{H}{1} deficiency is usually attributed to the process of ram 
pressure sweeping. In a similar way Williams \& Lynch (1991) detected a lower
than average gas content for the spiral members in four poor clusters, while  
Williams \& van Gorkom (1988); and Williams \& Rood (1987) detected a similar effect 
when analyzing a few compact groups.  Haynes (1981) found the presence of neutral 
hydrogen streams in 6 out of 15 groups of galaxies.  In the paper of Giuricin, 
Mardirossian \& Mezzetti (1985) they use galaxies in loose groups identified by 
Geller \& Huchra (1983) to examine the \ion{H}{1} content.  For a sample of 213 
spiral and irregular galaxies they do not observe evidence of gas removal nor do 
they find a variation of the \ion{H}{1} properties as a function of group compactness 
or even with the distance of the galaxies to the center of the groups.  The fact that 
Geller \& Huchra's (1983) groups are known to be plagued by interlopers may disguise 
some possible result in favor of \ion{H}{1} deficiency, particularly if this effect 
is not conspicuous. 
Those results indicate that in clusters gas removal does occur; what happens in less 
dense structures is still an open question. 
There is observational evidence at several wavelengths for the presence of an
intergalactic medium in loose groups, which could remove the gas from spiral 
galaxies (e.g., Dell'Antonio, Geller \& Fabricant 1994; Henry \etal 1995; 
Mulchaey \etal 1996).  Henry \etal (1995) claim that the x-ray luminosity and 
temperature functions may be considered as smooth extrapolations from that of rich 
clusters.  There are also radio observations with the VLA in 20cm by Burns 
\etal (1987) who report the presence of tailed radio galaxies, and attribute this 
fact to the existence of an intragroup medium.  

Besides the mechanism proposed above, tidal effects could be another possible 
process for gas removal which might be efficient in groups, since the velocity 
dispersions of those systems are low (typically $\approx$ 250 km s$^{-1}$) and close 
encounters of galaxies will last considerable time, allowing the external parts of 
the galaxies to be removed.  In fact, Davis \etal (1997) examining galaxies of 
loose groups using ROSAT, VLA and optical data concluded that both effects 
appear to be acting on the galaxies.  They also suggest that the stripping of gas 
by the intragroup medium is made more efficient after a gravitational encounter. 

Thus, we expect that the effect of gas removal may be present in some degree 
for density regimes such as those found in loose groups of galaxies.  
In this paper we generate a catalog of loose groups by means of objective criteria 
and the \ion{H}{1} content for the constituent galaxies is evaluated.  We also examine 
the possible role of the mechanisms for gas removal driven either by hydrodynamic  
or by gravitational forces, in a typical loose group environment. 
The selection criteria of the sample of galaxies as well as the group definition 
are described in section 2, the \ion{H}{1} content in group galaxies is analyzed 
in section 3. A brief conclusion is presented in section 4.

\section{THE SAMPLE AND LOOSE GROUP DEFINITION}

\subsection{The Galaxy Sample} 

In this work we analyze a magnitude limited sample of galaxies taken from 
the {\sl Catalog of Galaxies and Clusters of Galaxies} (Zwicky \etal 1961-68, 
hereafter CGCG), in the region of the sky defined by the intervals of declination 
+9\deg to +15\deg and right ascension of 8$^h$ to 18$^h$. A detailed description 
of the catalog, observations and data reduction will be given in a forthcoming 
paper, so only a brief description will be presented here.  
All the 2366 galaxies in this list up to $m_Z$=15.7 were visually inspected in 
overlays of the Palomar Observatory Sky Survey plates and had improved
measurements of coordinates, major and minor diameters, as well as morphological
types.  Whenever possible, multiple systems had these parameters measured and 
magnitudes estimated for individual members. For those CGCG galaxies also listed 
in the {\sl Uppsala General Catalog of Galaxies} (Nilson 1973, UGC) the 
morphological classification and diameter measurements were maintained as 
quoted in the UGC, unless we had a case of a split multiple system, when new
parameters were calculated.   Late-type galaxies (Sa and later) were selected to be 
observed with the 305 m Arecibo radiotelescope, while for the early-types as well as 
spirals which were not observed in 21 cm, we used the 2.15 m telescope of the Complejo 
Astronomico El Leoncito, San Juan, Argentina, the 1.6 m telescope of the 
Laborat\'orio Nacional de Astrof\'\i sica, Itajub\'a, Brazil, and the European 
Southern Observatory 1.52 m telescope\footnote{ The ESO 1.52 m telescope is operated 
under agreement between European Southern Observatory and Observat\'orio 
Nacional/CNPq - Brazil.}, La Silla, Chile.  
At the present time, the redshift completeness of this sample is better 
than 95\%.  The incompleteness is partly caused by the observational procedure 
used in the 21cm survey. Typically the search for 21cm emission was carried out 
in the interval  between 0 and $\approx$  16000 \kms using bands of about 6000 \kms. 
In the final period of the radio survey we carried out searches up to 25000 \kms 
and several galaxies not detected previously, proved to be in that new search  
interval of velocities.

\subsection{Determination of Loose Groups}

The algorithm adopted for the construction of the catalog of groups of galaxies  
is basically the one described by Huchra \& Geller (1982) with the 
improvements by Ramella \etal (1989) and by Maia, da Costa \& Latham (1987) to 
minimize the number of interlopers.  This percolation algorithm identifies groups 
of galaxies in a magnitude limited sample. 
A search for companions around galaxies is carried out taking into account 
projected separations satisfying 

$$D_{12} = 2 \sin(\theta_{12}/2) V / H_0  \leq D_L$$

and with line-of-sight velocity differences,

$$V_{12} = \mid V_1 - V_2 \mid  \leq V_L$$

In the above expressions $V = (V_1 + V_2) / 2$, $V_1$ and $V_2$ are the radial
velocities of the galaxies, and $\theta_{12}$ their angular separation.  The
quantities $D_L$ and $V_L$ are search parameters scaled according to the expressions below 
in order to take into account the variation in the sampling of the galaxy luminosity 
function, $\phi(M)$, with distance 

$$ D_L = D_0 R \quad ; \quad  V_L = V_0 R$$

where

$$ R =  \Biggl[\int_{-\infty}^{M_{12}} \Phi (M) dM \biggm / \int_{-\infty}^{M_{lim}} \Phi (M) dM
\Biggr]^{-1/3} $$

$$ M_{lim} = m_{lim} -25 - 5\log(V_f / H_0) $$

$$ M_{12} = m_{lim} -25 -5\log[(V_1 + V_2)/2 H_0] $$

$D_0$ is the selection parameter at a fixed fiducial radial velocity, $V_f$. 
$V_L$ is scaled in the same way as$D_L$.  For the present work, the adopted values 
for $D_0$ and $V_0$ are 0.223 Mpc and 350 \kms\ respectively; the apparent magnitude, 
$m_{lim} = 15.7$, and $H_o = 100$ \kms Mpc$^{-1}$. The groups present 
a surrounding density contrast  ($\delta\rho/\rho$)  relative to  the mean density 
of galaxies of 80 for a Schechter (1976) luminosity function calculated for the 
entire sample, parameterized by $\phi^*=0.025$ galaxies mag$^{-1}$ Mpc$^{-3}$, 
$M^*_{B(0)} =-19.34$ and $\alpha=-1.38$.

Although the values we find are steeper them those found in some recent surveys 
(da Costa \etal 1994, Marzke \etal 1994), they are consistent with those of Ramella 
\etal (1989) and are very likely to be affected by the presence of the Virgo cluster. 
A total of 73 groups with 4 or more members and mean velocities 
$\le$ 12,000 \kms\ were identified. 
The global physical parameters as well as the presence of interlopers do 
not depend strongly on redshift (see Ramella \etal 1989 for a more 
complete discussion).  
The physical parameters of the identified groups, are presented in Table 1 
where we list the median and upper and lower quartile values for the velocity 
dispersion, $\sigma_v$; virial mass, $M$; crossing time, $t_c$; virial radius, 
$R_h$; and the mean projected separation, $R_p$.  For a few parameters like number 
of member galaxies, $N_{mem}$, $t_c$, $ log\ M$ and $R_p$ we display in Figure 1 
histograms of the respective distributions.  The results from both Table 1 and 
Figure 1 suggest that the identified groups have average properties similar to 
those identified by Ramella \etal (1989).  The smaller crossing times and the short 
tail distribution of higher mass groups reflect the low contamination by interlopers.

\section{THE NEUTRAL HYDROGEN CONTENT OF GALAXIES IN LOOSE GROUPS}

A subsample called ``Group'' containing spiral galaxies with 21 cm information, 
members of the groups generated above is used to have the  \ion{H}{1} content of 
their galaxies estimated.  We did not include in the analysis the group identified 
as part of the Virgo cluster because our intent is to examine the neutral Hydrogen 
content in structures with density regimes of groups. The galaxies which were not 
assigned to groups formed a subsample called ``Isolated'' that is used as a control 
sample. Isolated galaxies whose separation from the survey boundaries were smaller 
than the projected search radius appropriate for their radial velocity were removed 
from the ``Isolated'' sample.  
For both subsamples, only galaxies in the range of absolute magnitudes 
$-21 \le M_{Z} \le -17$, and radial velocities $\le 12,000$ \kms were considered 
in this analysis. The distribution of velocities for the galaxies in both 
samples is more or less the same, excluding the possible bias of a particular 
sample to be made up of only distant or nearby objects.  Furthermore we have examined
whether more massive galaxies could be more susceptible to loose their gas by
examining the distribution of the estimators with redshift for each morphological 
type and no such trend was detected.  

Two estimators of the \ion{H}{1} content are used.  The first one, is the 
{\it Pseudo} \ion{H}{1} surface density, $\Sigma_{{\scriptscriptstyle HI}}$, 
in M$_\odot$pc$^{-2}$, which is the total \ion{H}{1} mass, 
$M_{\scriptscriptstyle HI}$, in solar units divided by the optical area of the 
galaxy, defined as 
 $$ \Sigma_{{\scriptscriptstyle HI}} = M_{\scriptscriptstyle HI} / 
(\pi/4)D_B^2) \ ,$$ 
where 
$$M_{\scriptscriptstyle HI} = 2.356 \times 10^5 \ d^2 F_c \ ,$$
and $D_B$ is the optical diameter in kpc, $d$ is the distance in Mpc, and $F_c$ 
is the 21cm line flux corrected for 
galaxy self-absorption, in Jy/\kms as described by Haynes \& Giovanelli (1984).  
The second estimator is the \ion{H}{1} mass-to-light ratio,  
$M_{\scriptscriptstyle HI} / L_{\scriptscriptstyle B}$, in solar units, assuming a 
solar photographic absolute magnitude of +5.37.  This estimator removes the 
dependence of the \ion{H}{1} content on the blue galaxy luminosity, as well as 
reducing the scatter about the mean values than $M_{\scriptscriptstyle HI}$ alone 
(Giovanelli \& Haynes 1988).  By compiling a larger sample and using additional
procedures, such as the one proposed by Solanes, Giovanelli \& Haynes (1996) which 
takes into account a complete \ion{H}{1} flux-limited data set, another methodology 
could be used. For the purpose of this paper we adopt the two estimators described 
above to perform the analysis of the \ion{H}{1} content.

Table 2 presents the statistical results of the 
$\Sigma_{{\scriptscriptstyle HI}}$ distribution for both samples. Column (1) lists  
the morphological type, column (2) the number of galaxies of a given morphological 
type, $N$; column (3) the mean, $\bar \Sigma_{HI}$; column (4) the median, 
$(\Sigma_{\scriptscriptstyle HI})_{med}$; column (5) the standard deviation, 
$\sigma$; columns (6) and (7) the lower ($LQ$) and  upper ($UQ$) quartiles of the 
distributions for ``Isolated'' galaxies. Columns (8) to (13), the same parameters 
as  columns (2) to (7) for ``Group'' galaxies. 
The Kolmogorov-Smirnov test (KS), was applied to both subsamples for each
morphological type, in order to determine whether they could all come from the 
same parent population.  Column (14) of Table 2 contains the probability, 
($P_{KS}$), of this hypothesis occurring by chance.   

In general, the statistical results presented in Table 2 show no significant 
dependence of $\Sigma_{{\scriptscriptstyle HI}}$ on the environment. 
Only the Sa and Sb galaxies in loose groups present some trend towards \ion{H}{1} 
deficiency, while the intermediate sample of Sab galaxies presents no such effect.  
The results are summarized in figure 2, where we show the 
$(\Sigma_{{\scriptscriptstyle HI}})_{med}$ (dots) with respective $LQ$ and 
$UQ$ (bars) of the $\Sigma_{{\scriptscriptstyle HI}}$ distributions for each 
morphological type.  

The statistics for the $M_{\scriptscriptstyle HI} / L_{\scriptscriptstyle B}$ 
estimator are presented in Table 3 and Figure 3.  Here the evidence for the 
presence of \ion{H}{1} depleted galaxies in groups is slightly stronger.  
Again, the earlier 
galaxies seem to be more affected by gas removal in the loose-group environment.  
The Sa-b galaxies with this estimator now show a tendency of being affected by the 
environment in contrast to the results obtained with 
$\Sigma_{{\scriptscriptstyle HI}}$.  For the Sbs the evidence weakens significantly.  
The fact that the intrinsic scatter of the estimators for each morphological type is 
usually high, may produce these fluctuations on the results making it difficult to 
detect an unambiguously clear evidence towards \ion{H}{1} deficiency, if present 
at all.  Also, the large dispersion of the results may be partially caused by the 
fact that the sample is not \ion{H}{1} flux limited.

 A comparison between our results for  
 $M_{\scriptscriptstyle HI} / L_{\scriptscriptstyle B}$ with others published in 
the literature is displayed in Figure 4, which shows the average values 
and standard deviations of 
 $M_{\scriptscriptstyle HI} / L_{\scriptscriptstyle B}$ for galaxies 
of the Virgo cluster (``Cluster'') obtained by Huchtmeier \& Richter (1989),  
for ``Group'' and ``Isolated'' galaxies as defined in this work, and for 
the isolated sample (Iso2) of Haynes \& Giovanelli (1984).  The binning of Sa with Sa-b 
has been chosen to allow us to compare our results with those of the authors above. In
Figure 4  we find that the control sample we have used presents an agreement with 
the Iso2.  The smaller scatter in our "Isolated" sample is due to the restrictions 
in the absolute magnitude interval we have applied.  There is also a clear trend 
towards the \ion{H}{1} deficiency for cluster galaxies but for ``Group'' galaxies, 
this effect is marginal.   

 An alternative mechanism for gas removal would be through tidal interactions.  To examine 
this possibility, we have divided the ``Group'' sample according to the values of 
$\sigma_v$ of their respective groups. 
The $M_{\scriptscriptstyle HI} / L_{\scriptscriptstyle B}$ is examined for each 
subsample, which present values of $\sigma_v$  smaller and higher than the mean 
value for the entire group sample which is 231 \kms .  The results, displayed in 
Figure 5, do not present any systematic behavior with $\sigma_v$.  In fact the Sa 
galaxies which we believe to show the strongest evidence of gas depletion, based on 
KS test, present the opposite behavior of what would be expected if tidal
interactions would be the case.  Therefore, this result tends to support the
interpretation that early spirals have gas depletion by ram pressure stripping. 
However, it is intriguing that Scs when examined as a function of $\sigma_v$ show 
a hint that systems with low values of $\sigma_v$ are significantly more 
depleted than those with higher values.  This seems to be in agreement with the 
results of Davis \etal (1997).

\section{SUMMARY}

We have examined the dependence of the \ion{H}{1} content of spiral galaxies 
with local environment.  For this purpose we identified loose-groups of galaxies 
by means of objective criteria.  This procedure also allowed us to define a control 
sample of isolated galaxies.  Two estimators were used to measure the \ion{H}{1} 
content of galaxies: $\Sigma_{{\scriptscriptstyle HI}}$ and 
$M_{\scriptscriptstyle HI} / L_{\scriptscriptstyle B}$; both estimators show 
no clear tendency of spiral galaxies in loose-groups having a lower amount of 
neutral Hydrogen, although there may be a hint for the earliest spirals, but 
because of the small number of objects involved we cannot claim these results 
as being statistically significant.  
We have also evaluated whether the \ion{H}{1} content of spiral galaxies shows 
any correlation with the group velocity dispersion.  Our results are inconclusive.  
Ram pressure may be a probable mechanism for producing the gas removal of 
galaxies in groups.  However, the contribution to the gas stripping given by  
gravitational forces produced in close encounters cannot be discarded.  
 By compiling a larger sample and using additional procedures, such as the 
one proposed by Solanes, Giovanelli \& Haynes (1996), it might be possible to 
investigate the importance of hydrodynamical and gravitational 
forces to remove gas from galaxies of loose-groups.

MAGM would like to thank the Astronomy Department of Cornell University for the 
hospitality during the course of part of this work.
This research was supported in part by the CNPq grants 204357/88.6, 301366/86-1 (MAGM), 
and 301364/81-9 (CNAW).  The resources of NASA Extragalactic Database (NED) were used 
during the course of this work.

\newpage

\figcaption[Maia.f1hi.ps]{ Histograms for different group parameters: number 
of member galaxies ($N_{mem}$), the crossing time ($t_c$) in units of Hubble 
time ($t_0$), The logarithm of the group mass ($log M$) in units of $M_{\sun}$, 
and the virial radius ($R_h$) in Mpc. \label{fig1}}

\figcaption[Maia.f2hi.ps]{Medians, upper and lower quartiles of the 
$\Sigma_{{\scriptscriptstyle HI}}$ distributions discriminating galaxies 
between the different morphological types. \label{fig2}}

\figcaption[Maia.f3hi.ps]{Medians, upper and lower quartiles of the 
$M_{\scriptscriptstyle HI} / L_{\scriptscriptstyle B}$ distributions 
discriminating galaxies between the different morphological types. 
\label{fig3}}

\figcaption[Maia.f4hi.ps]{Mean and standard deviations for  
$M_{\scriptscriptstyle HI} / L_{\scriptscriptstyle B}$ distributions. 
The ``Cluster'' sample refers to Virgo cluster galaxies by Huchtmeier \& 
Richter (1989); ``Group'' and  ``Isolated'' as defined in this work, 
and ``Iso2'' for the isolated sample of Haynes \& Giovanelli (1984). 
\label{fig4}}

\figcaption[Maia.f5hi.ps]{Medians, upper and lower quartiles of the 
$M_{\scriptscriptstyle HI} / L_{\scriptscriptstyle B}$ distributions 
for subsamples of galaxies which belong to groups with values of 
 $\sigma_v$ smaller and larger than the median value 231 \kms of the 
entire sample of groups. 
\label{fig5}}


\begin{references}
\normalsize

\reference {} Burns, J.O. \etal 1987, \aj 94, 587

\reference {} da Costa, L. N., et al. 1994, ApJ, 424, L1

\reference {} Davis, M., \& Geller, M.J. 1976, \apj\ 208, 13 

\reference {} Davis, D.S., Kell, W.C., Mulchaey, J.S., \& Henning, P.A. 1997, 
preprint astro-ph9705220

\reference {} Dell'Antonio, I.P. Geller, M.J., \& Fabricant, D.G. 1994, \aj 107, 427

\reference {} Dressler, A. 1980, \apj\ 236, 351

\reference {} Geller, M.J., \& Huchra, J.P. 1983, \apjs 52, 61

\reference {} Giovanelli, R., \& Haynes, M.P. 1988, in ``Galactic and Extragalactic 
Radio Astronomy'', ed. G.L. Verschuur \& K.I. Kellermann (Springer, New York), 522 
 
\reference {} Giuricin, G., Mardirossian, F., \& Mezzetti, M. 1985, \aap 146, 317 

\reference {} Haynes, M.P. 1981, \aj\ 86, 1126

\reference {} Haynes, M.P., \& Giovanelli, R. 1984, \aj\ 89, 758

\reference {} Haynes, M.P., \& Giovanelli, R. 1986, \apj\ 306, 466

\reference {} Henry, J.P. \etal 1995, \apj 449, 422

\reference {} Huchra, J.P., \& Geller, M.J. 1982, \apj\ 257, 423

\reference {} Huchtmeier, W.K., \& Richter, O.-G. 1989, \aap\ 210, 1

\reference {} Maia, M.A.G., \& da Costa, L.N. 1990, \apj\ 352, 457

\reference {} Maia, M.A.G., da Costa, L.N., \& Latham, D.W. 1989, \apjs\ 69, 809 

\reference {} Maia, M.A.G., Pastoriza, M.G., Bica, E., \& Dottori, H. 1994, 
\apjs\ 93, 425

\reference {} Magri, C., Haynes, M.P., Forman, W., Jones, C., \& Giovanelli, R. 1988,
\apj\ 333, 136

\reference {}  Markze, R. O., Huchra, J. P., \& Geller, M. J.
1994, ApJ 428, 43 

\reference {} Mulchaey, J.S., Davis, D.S., Mushotzky, R.F., \& Burstein, D. 1996, 
\apj\ 456, 80 

\reference {} Nilson, P. 1973, Uppsala General Catalog of Galaxies, (Uppsala 
Astron. Obs.) (UGC)

\reference {} Postman, M., \& Geller, M.J. 1984, \apj\ 281, 95

\reference {} Ramella, M., Geller, M.J., \& Huchra, J.P. 1989, \apj\ 344, 57

\reference {} Schechter, P.L. 1976, \apj\ 203, 297

\reference {} Scodeggio, M., \& Gavazzi, G. 1993, \apj\ 409, 110

\reference {} Solanes, J.M., Giovanelli, R., \& Haynes, M.P. 1996, \apj\ 461, 609

\reference {} Williams, B.A., \& Lynch, J.R. 1991, \aj\ 101, 1969

\reference {} Williams, B.A., \& Rood, H.J. 1987, \apjs\ 63, 265 

\reference {} Williams, B.A., \& van Gorkom, J.H. 1988, \aj\ 95, 352

\reference {} Zwicky, F., Herzog, E., Karpowicz, M., \& Kowal, C. 1961-1968. 
Catalog of Galaxies and Clusters of Galaxies, 6 vols. (Pasadena: California 
Institute of Technology) (CGCG)

\end{references}
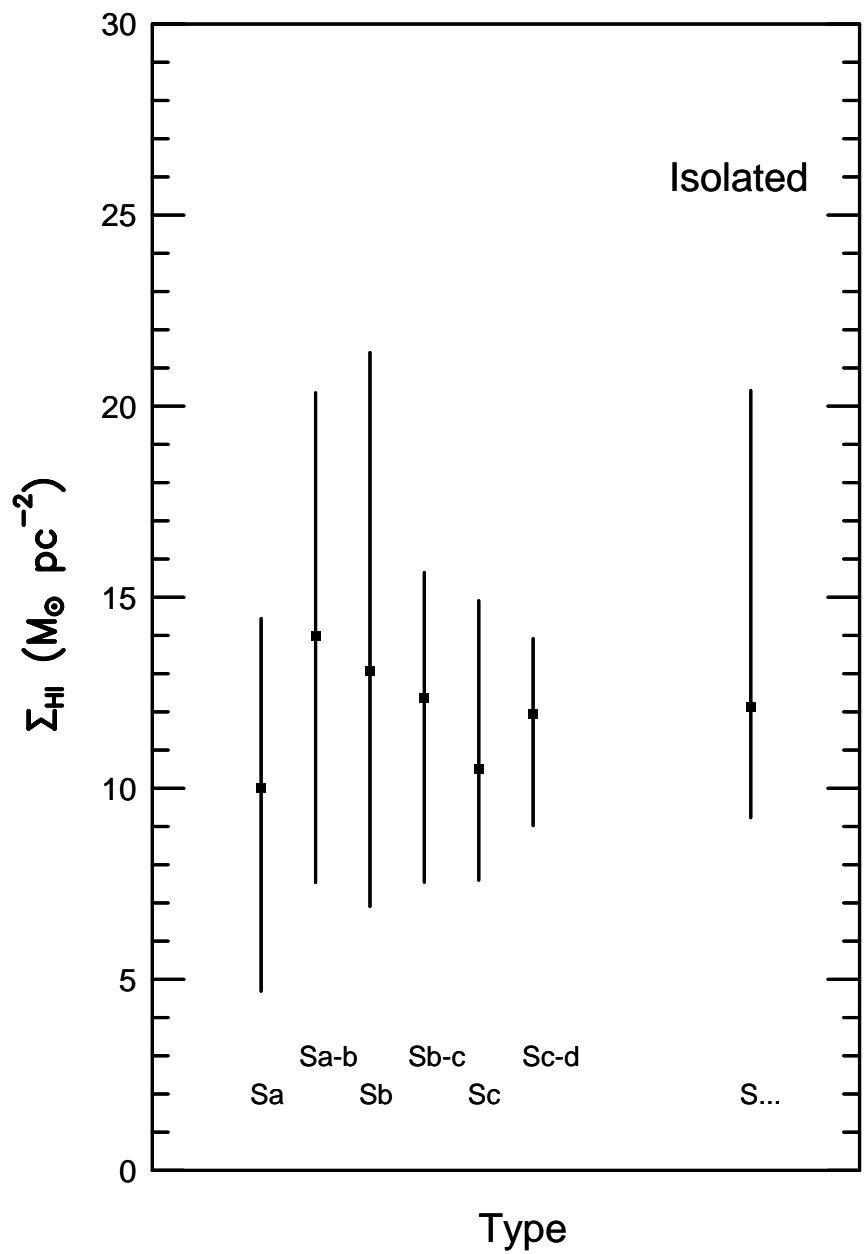
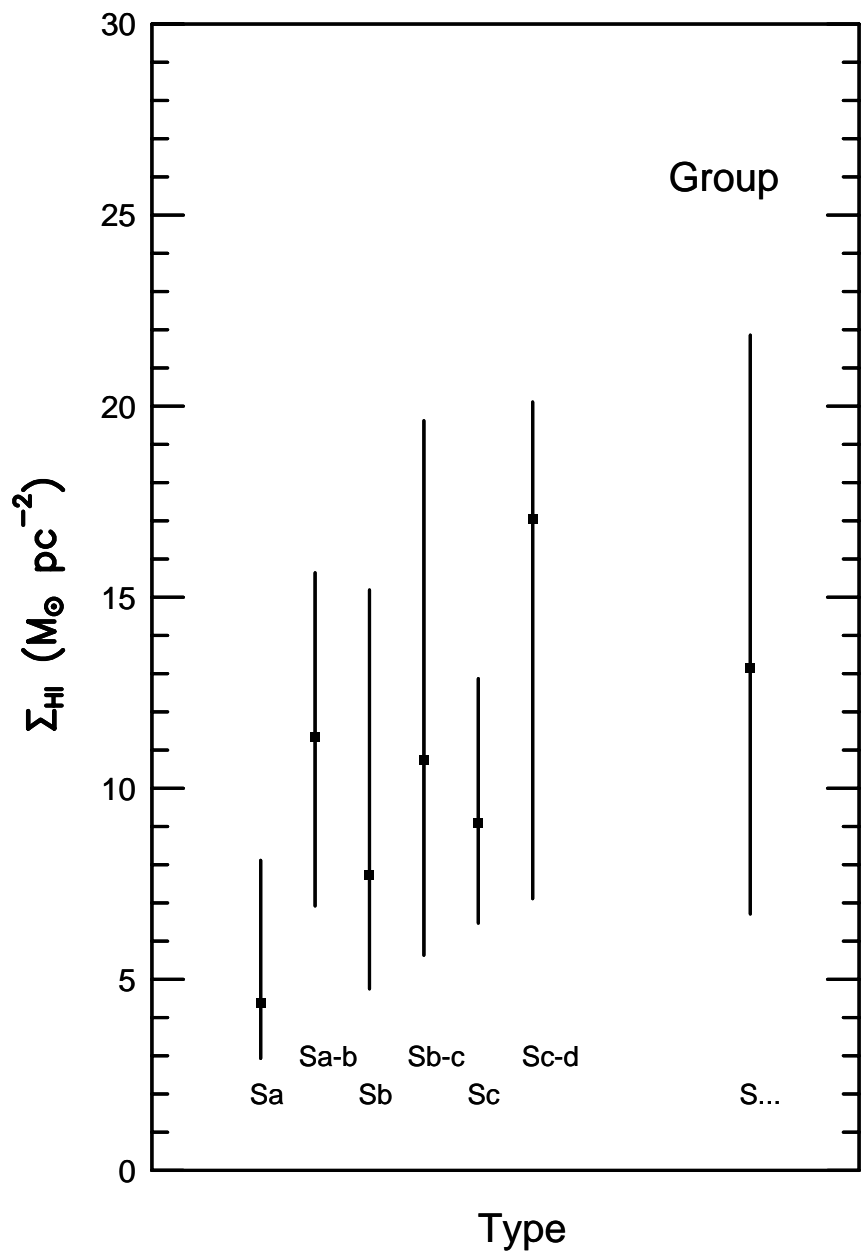
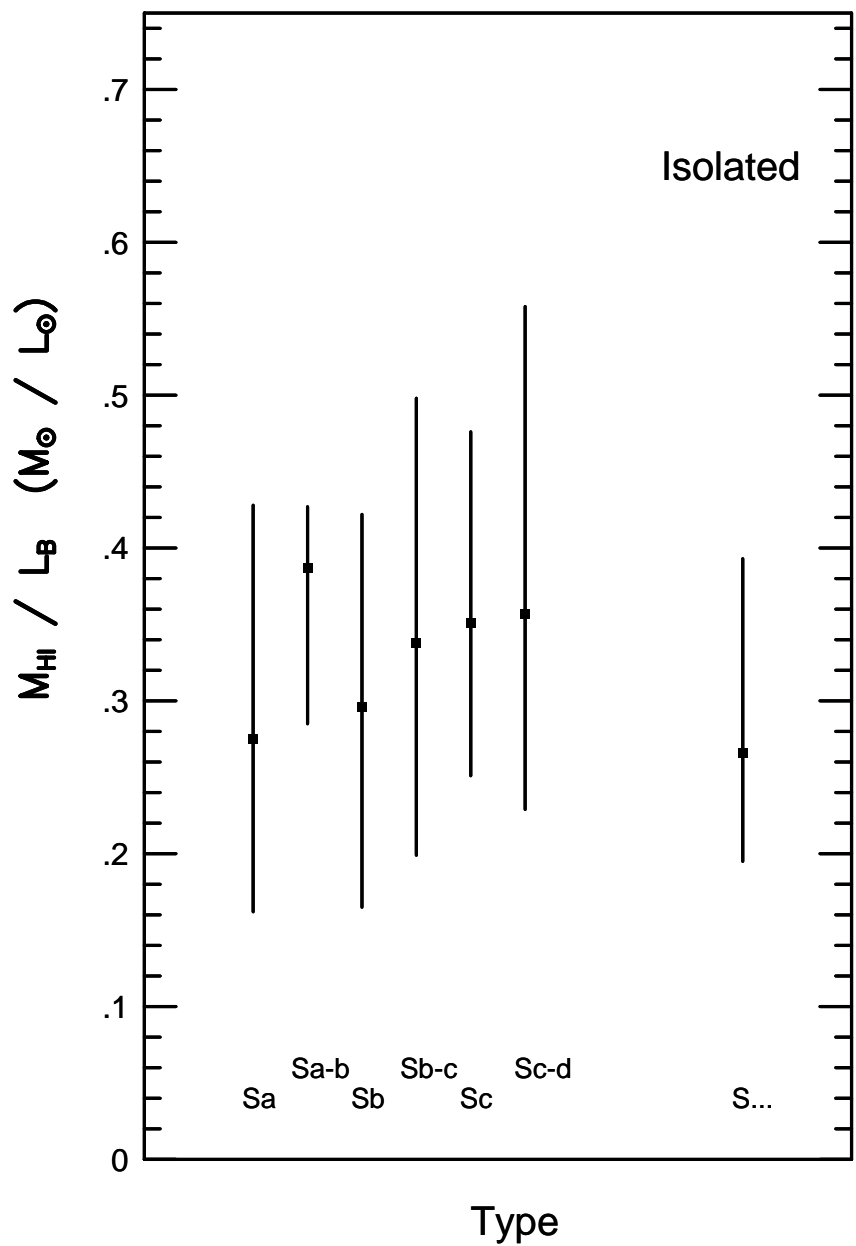
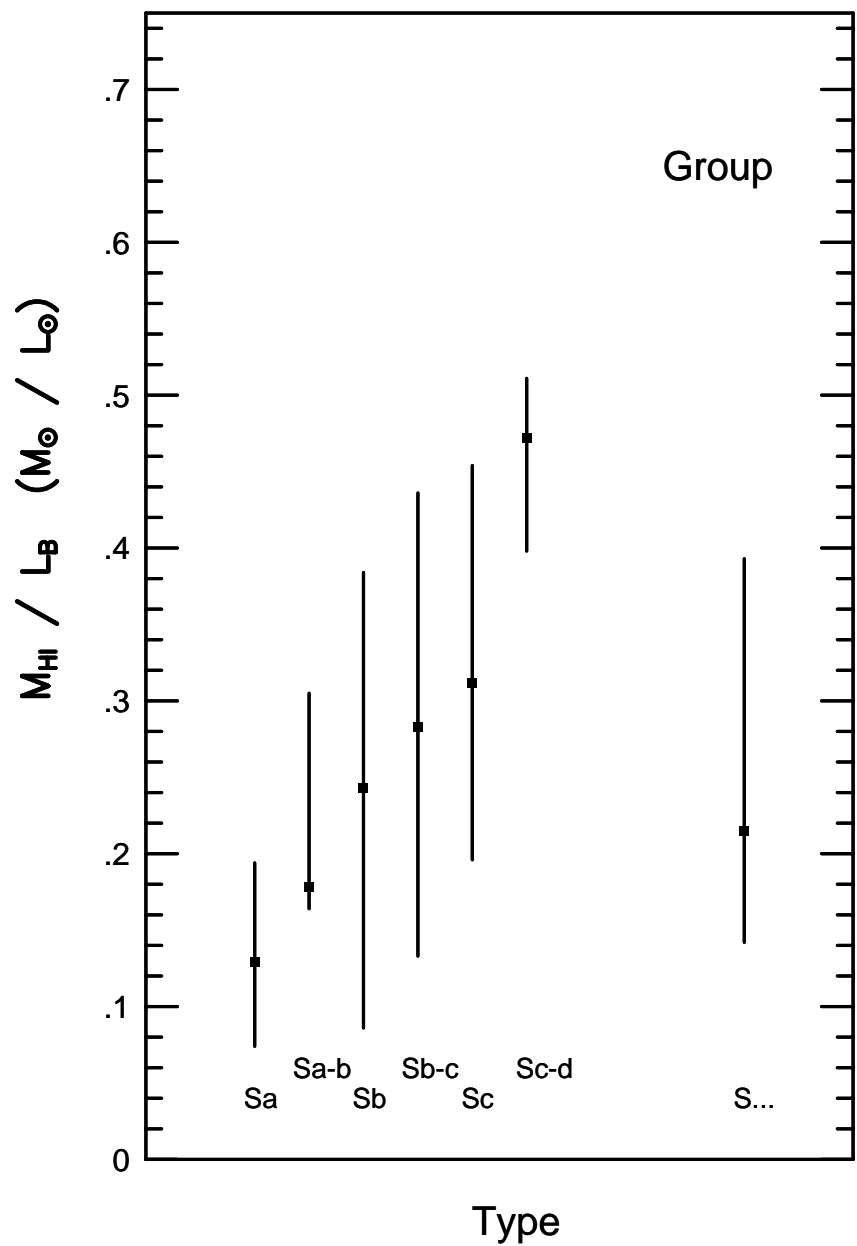
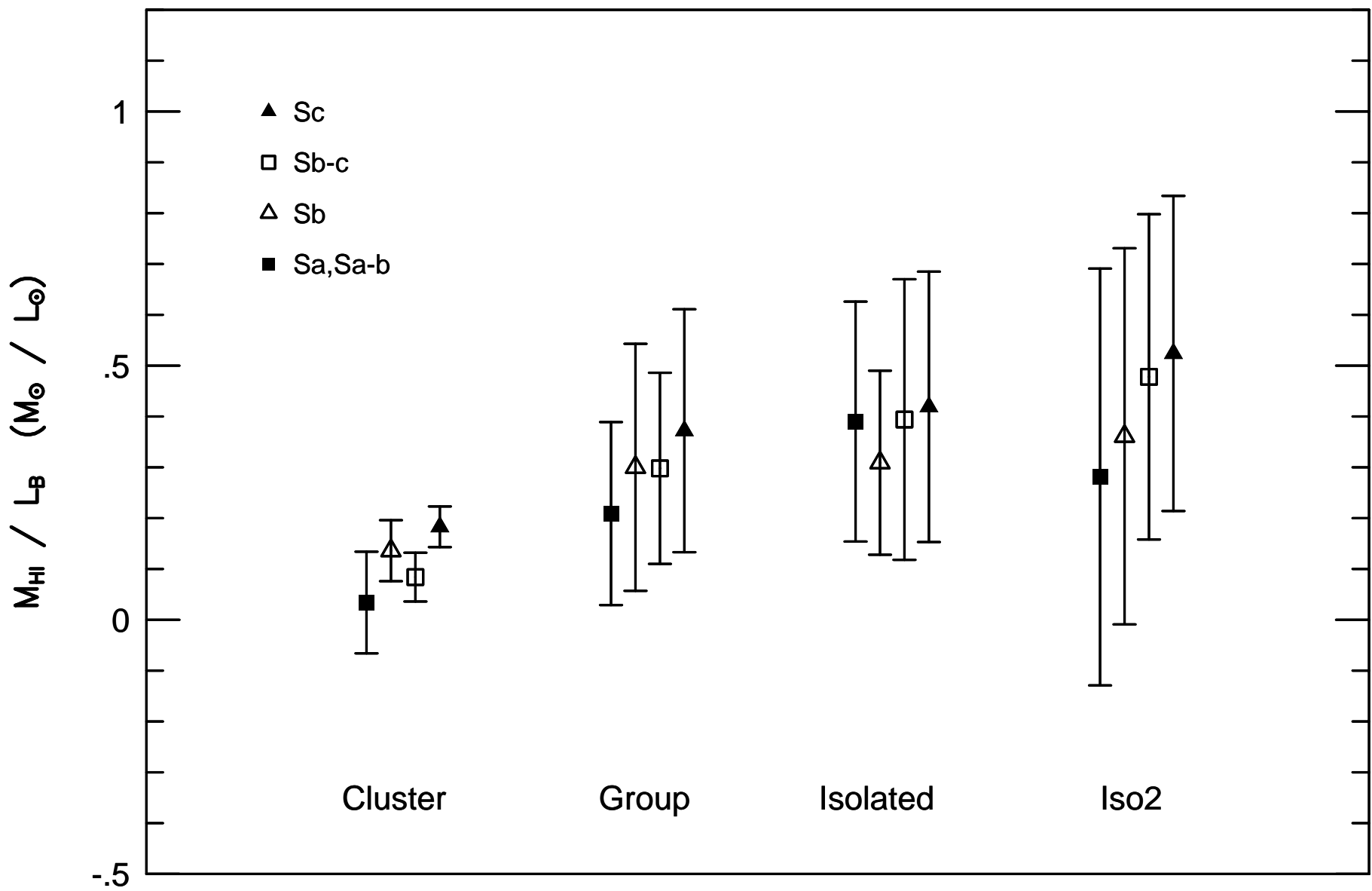
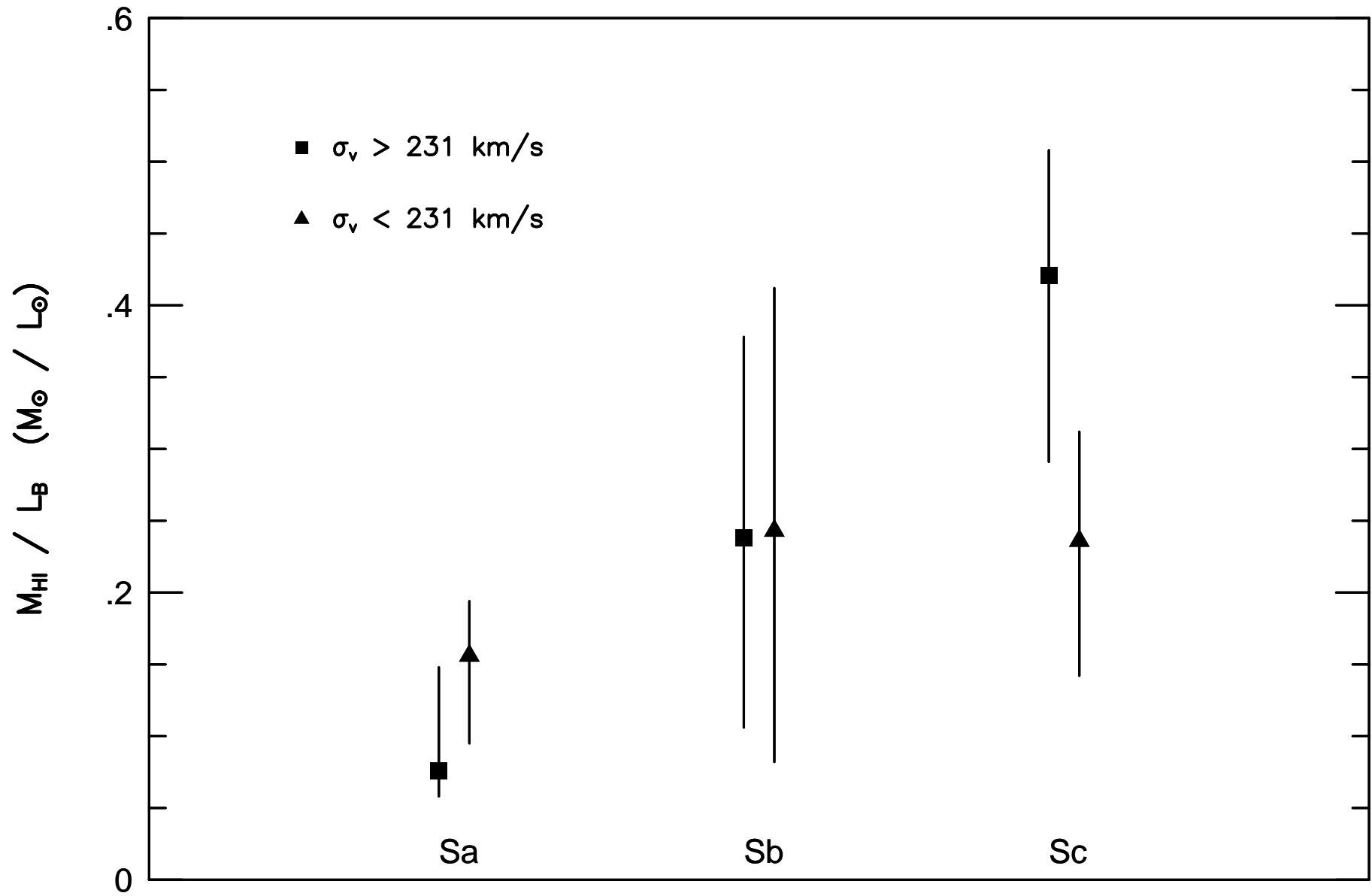
\end{document}